\documentclass[journal=cmatex,manuscript=article,layout=twocolumn]{achemso}
\usepackage[english]{babel}
\usepackage[utf8]{inputenc}
\usepackage{amsmath}
\usepackage{graphicx}
\usepackage{float}
\usepackage{enumitem}
\setlist{noitemsep}
\usepackage[caption=false]{subfig}
\usepackage{siunitx}

\graphicspath{{./figures/}}

\newcommand{\full}{$\mathrm{C_{60}}$}
\newcommand{\irppy}{$\mathrm{Ir(ppy)_3}$}
\newcommand{\etal}{\textit{et al.}}
\newcommand{\ie}{\textit{i.e.}}
\newcommand{\eg}{\textit{e.g.}}
\title{Atom Probe Tomography of Molecular Organic Materials: Sub-Dalton Nanometer-Scale Quantification}
\author{Andrew P. Proudian}
\affiliation{Department of Physics, Colorado School of Mines}
\author{Matthew B. Jaskot}
\affiliation{Materials Science Program, Colorado School of Mines}
\author{David R. Diercks}
\affiliation{Materials Science Program, Colorado School of Mines}
\alsoaffiliation{Department of Metallurgical and Materials Engineering,
Colorado School of Mines}
\author{Brian P. Gorman}
\affiliation{Materials Science Program, Colorado School of Mines}
\alsoaffiliation{Department of Metallurgical and Materials Engineering,
Colorado School of Mines}
\author{Jeramy D. Zimmerman}
\affiliation{Department of Physics, Colorado School of Mines}
\alsoaffiliation{Materials Science Program, Colorado School of Mines}
\email{jdzimmer@mines.edu}
\date{\today}

\begin{document}

\maketitle

\begin{abstract}
In this paper, we demonstrate that atom probe tomography (APT) can be applied to small-molecule organic materials. We show that APT can provide an unprecedented combination of mass resolution of $<\SI{1}{\dalton}$, spatial resolution of $\sim\SI{0.3}{\nm}$ in z and $\sim\SI{1}{\nm}$ in x-y, and analytic sensitivity of $\sim\SI{50}{ppm}$ with no evidence of molecular fragmentation. We discuss two systems that demonstrate the power of APT to uncover structure-property relationships in organic systems that have proven extremely difficult to probe using existing techniques:
(1) a previously published model organic photovoltaic system in which we show a chemical reaction occurs at the heterointerface;
and (2) a model organic light-emitting diode system in which we show molecular segregation occurs in the emissive layer bulk.
These examples illustrate the power of APT to enable new insights into organic molecular materials.
\end{abstract}

\section{Introduction}
Modern electronics now include a wide variety of organic devices, such as organic photovoltaics (OPVs), organic thin film transistors (OTFTs), and organic light-emitting diodes (OLEDs). Compared to their inorganic counterparts, organic electronic devices enjoy a number of advantages, including low-cost room-temperature deposition, easy patterning at relevant length scales (\eg\ display pixels), mechanical flexibility, and application-specific tunability; the broad commercial success of OLEDs in the past few years provides a clear example of this.%
\cite{Gaspar2015}
However, to fully realize these advantages requires more detailed knowledge of the structure-property relationships of these devices.
As we will show, atom probe tomography (APT) is a valuable tool for advancing our knowledge of these molecular systems.

The fundamental physics of inorganic semiconductors has been well-established for over half a century, though engineering challenges remain.
In contrast, there are still many developments to be made in the fundamental physics of organic semiconducting materials.%
\cite{Kippelen2009,Giebink2010,Lin2018}
A major difference for organic electronics---which makes them more difficult to describe theoretically---is the strong influence of small changes to morphology on device performance and reliability.%
\cite{Kippelen2009,Maturova2009,Brady2011}
Many methods have been explored to determine how microstructure changes with material, deposition, and processing conditions;%
\cite{Jorgensen2008,Groves2010,Smith2011,Brady2011,Leong2013,%
Neuhold2013,Vandewal2013,Fu2014,Lan2016,Vinokur2017,Lin2018}
in addition, morphological changes and molecular degradation impact the performance and lifetime of these devices.%
\cite{DAndrade2003,Smith2011,DeMoraes2012,Cho2016,%
FragaDominguez2017,Mateker2017,Dong2017,Bangsund2018}
These studies have led to improvements in performance, but have largely been driven by empirical investigation rather than material theory.%
\cite{Jorgensen2008,Groves2010,Brady2011}
Improved structural characterizations of these devices in three dimensions will lead to better physical theories of organic electronics, shortening development times.%
\cite{Jorgensen2008,Maturova2009,Groves2010,Brady2011,Chen2012,%
Pfannmoller2013,Roehling2013}

Because their properties are strongly dependent on morphology, organic films require detailed nanoscale characterization.%
\cite{Brady2011,Pfannmoller2013}
For inorganic systems, a wide variety of tools are available for high-resolution imaging and tomography.%
\cite{Tennyson2017}
Unfortunately, many of the techniques are challenging to apply to  organic systems, as they are sensitive to ion, electron, and X-ray irradiation, and many techniques are hampered by weak scattering contrast between materials.%
\cite{Chen2012,Neuhold2012,Schindler2012,Love2013,Pfannmoller2013,%
Orthacker2014}
This does not rule out these techniques, but limits the kind and quality of data they can yield.

Transmission electron microscopy (TEM) and its derivatives (\eg\ energy-filtered TEM (EF-TEM) high-angle annular dark-field scanning TEM (HAADF-STEM)) have enabled ground-breaking studies of organic nanostructures because of their excellent spatial resolution---with some instruments capable of resolving sub-nanometer features---which greatly contributed to the development of organic electronic devices.%
\cite{Jorgensen2008,Maturova2009,Brady2011,Chen2012,%
Love2013,Pfannmoller2013,Deng2016}
Unfortunately, difficulty with contrast in these systems can make it challenging to definitively identify composition.%
\cite{Chen2012,Schindler2012,Pfannmoller2013,Roehling2013}
Scattering methods are also a common tool for probing organic film structure, and have been instrumental in developing our understanding of organic morphologies; however, due to the low scattering contrast or requirement for a tunable source, they must often be performed at beamlines or user facilities.%
\cite{Jorgensen2008,Vickers2010,Brady2011,Smith2011,Chen2012,%
Love2013,Neuhold2013,Pfannmoller2013,Deng2016}
A variety of other methods have been used to probe nanostructure as well,
such as atomic force microscopy,%
\cite{Shaheen2001,VanDuren2004,McNeill2005,Jorgensen2008,Maturova2009,%
Smith2011,Chen2012,Neuhold2013,Pfannmoller2013,Deng2016}
various optical techniques (\eg\ UV-vis/IR spectroscopy),%
\cite{Jorgensen2008,Brady2011}
X-rays (\eg\ near edge X-ray absorption fine structure spectroscopy),%
\cite{Jorgensen2008,Brady2011,Chen2012,Pfannmoller2013}
and secondary-ion mass spectroscopy (SIMS).%
\cite{VanDuren2004,Jorgensen2008,Brady2011,Chen2012,%
Popczun2017,Vinokur2017}
SIMS is the closest to a single measurement of chemical and morphological information that is widely used by the community, but still cannot answer many questions about system morphology because of its limited spatial resolution and fragmentation of molecular species.

The immense effort to develop and adopt new tools has enabled innumerable advances in the field, but many questions remain that are difficult to explore with existing techniques. Existing imaging techniques generally have insufficient resolution or chemical contrast to characterize the nanostructure of the material at adequate levels to inform better physical theories of organic electronics; measurements typically provide either high spatial or sensitive chemical information---but rarely both---requiring indirect correlation of information from several different techniques.%
\cite{Chen2012,Pfannmoller2013}

APT, which combines both high spatial resolution ($<\SI{1}{\nm}$) and high analytical sensitivity ($<\SI{100}{ppm}$) in three dimensions, simultaneously measures chemical and spatial information with high precision.%
\cite{Kelly2007,Kelly2012}
This ability can help provide the structural information needed to inform the next generation of physical theories for organic electronics.

In APT, a sample prepared with a sub-micrometer radius of curvature is held at cryogenic temperatures under high bias, generating a large local electric field. A voltage or laser pulse of adequate intensity causes an atom or molecule on the surface to field evaporate. The electric field accelerates this ion towards a two-dimensional position- and time-sensitive detector.
%;we note that the efficiency of this detector is up to \SI{80}{\percent} in new instruments,%
%\cite{Ulfig2017}
%which is quite high for any ion counting technique.%
%\cite{Geiser2007}
This process repeats until the desired thickness of sample has evaporated. The time-of-flight of each ion gives its mass-to-charge ratio, and its position on the detector allows the ion's location to be reconstructed in three dimensions.%
\cite{Bas1995,Kelly2007,Kelly2012}

Despite APT's excellent spatial and chemical resolution and the relative maturity of the technique, the body of APT analyses of organic systems is small.%
\cite{Machlin1975,Panitz1982,Panitz1983,Nishikawa1998,%
Nishikawa1986,Maruyama1987,Zhang2009,Prosa2010,Joester2012}
Most of these studies looked at polymers, which must fragment to field evaporate because of their polymeric structure.%
\cite{Nishikawa1986,Maruyama1987,Zhang2009,Prosa2010,Joester2012}
In 2012, Joester \etal\ examined a blend of poly(3-hexylthiophene-2,5-diyl) (P3HT) and \full;%
\cite{Joester2012}
as before, the P3HT polymer proved difficult to study due to uneven fragmentation, but the mass-spectrum had a clear \full\ signal, suggesting that small-molecule organic systems should be amenable to study with APT; this can be understood by considering that the strength of the intra-molecular (covalent) bonds in small-molecule organics is roughly 2-4 times that of  the inter-molecular (van der Waals) bonds.%
\cite{Chickos2003}

In this paper, we show that APT can have a mass resolution of $<\SI{1}{\dalton}$, spatial resolution of $\sim\SI{0.3}{\nm}$ in z and $\sim\SI{1}{\nm}$ in x-y, and an analytic sensitivity of $\sim\SI{50}{ppm}$ for these materials with no evidence of molecular fragmentation. This capability enables new insights into structure-property relationships at the nanoscale---both at interfaces and in the bulk---which we demonstrate in an OPV%
\cite{Proudian2016}
and an OLED system. This information can help drive forward the development and deployment of molecular organic electronics.

\section{Methods}
\subsection{Sample Preparation}
Thin films were prepared by vacuum thermal evaporation at room temperature, using rates of $\SI{1}{\angstrom\per\second}$ for the primary constituent; the film thickness and doping ratios were controlled by the deposition rate, monitored by a quartz crystal oscillator.

4,4'-bis(N-carbazolyl)-1,1'-biphenyl (CBP) and tris[2-phenylpyridinato-C2,N]iridium(III) (\irppy) were from Luminescence Technology Corporation; CBP was used both as received and purified by vacuum gradient sublimation before use, while \irppy\ was used as received. Sublimed grade \full\ from MER Corp. was purified by vacuum gradient sublimation before use.

Because of the sensitivity of organic films to electron radiation, we deposit films directly on a curved Si or W tip; we note that a W tip was used in our OPV study, while the rest of the data presented here used Si tips.%
\cite{Proudian2016}
While direct deposition is common practice for the organic electronics community, this method of sample preparation is very uncommon within the APT community, which mostly prepares tips by cutting samples out from a flat film using a focused ion beam (FIB).%
\cite{Kelly2007}
Because of the low evaporation field of molecular organic materials, we can use a tip with a comparatively large radius of curvature for APT ($R =$ \SIrange{250}{500}{\nm}) which approximates a flat surface locally. The Supporting Information provides an SEM image of a Si tip in Figure S1 and a reconstruction cross-section showing the lenticular shape of the reconstructed film in Figure S2.

\subsection{Data Acquisition and Analysis}
APT data were taken using a CAMECA LEAP 4000X Si operated in laser pulse mode using a \SI{355}{\nm} laser; the multiple hit values are typically below \SI{15}{\percent}. The reconstructions were performed using CAMECA IVAS software (3.6.14).

Because the organic molecules have much larger masses than the typical ions observed in APT, the laser pulse repetition rate must be significantly slower than usual to allow for the required time of flight to the detector. Beyond that adjustment and the larger sample radius discussed above, no special accommodations are needed and good running parameters are within normal ranges for the instrument. The chosen parameter values were optimized to minimize multiple hit events and maximize mass spectral resolution; in general, higher laser pulse energies increase the multiple hit rate and decrease mass spectral resolution, and higher temperatures decrease mass spectral resolution as well. The analysis window seems relatively broad, as we did not identify any particularly sensitive parameters. The values presented below are typical for our specimens, and we list a table of analysis parameters for various materials in the Supporting Information (Table S1).

The crystalline \full\ sample (analyzed in Figures \ref{fig:SDM}, S10, and S11) used a temperature set-point of \SI{30}{\kelvin}, a laser pulse energy of \SI{12}{\pico\joule}, an evaporation rate target of 1 ion per 100 pulses, a laser pulse frequency of \SI{250}{\kilo\hertz}, and a detector distance of \SI{90}{\mm}. It was reconstructed using voltage evolution with an image compression factor (ICF) of 1.60, a detector efficiency of 0.55, and a tip radius of \SI{315}{\nm}. The spatial distribution map (SDM) was generated using \SI{0.05}{\nm} z cuts to give the best peak quality.

The blended CBP:\irppy\ sample (Figures \ref{fig:impurity} \&\ \ref{fig:K}) used a temperature set-point of \SI{25}{\kelvin}, a laser pulse energy of \SI{5}{\pico\joule}, an evaporation rate target of 3 ions per 100 pulses, a laser pulse frequency of \SI{125}{\kilo\hertz}, and a detector distance of \SI{160}{\mm}. It was reconstructed using voltage evolution with an ICF of 2.00, a detector efficiency of 0.55, and a tip radius of \SI{350}{\nm}.

A representative voltage curve and detector hit map are provided in the Supporting Information (Figures S3 \& S4). Mass spectra analysis was performed using the MALDIquant package in R, using the Statistics-sensitive Non-linear Iterative Peak-clipping (SNIP) algorithm for background subtraction.%
\cite{Morhac2009}
Spatial statistical analysis was performed using the spatstat package in R, or extensions available at \url{https://github.com/aproudian2/rapt}.

\section{Results and Discussion}
\begin{figure} % Line drawings of molecules
\centering
\includegraphics[width=3.3in]{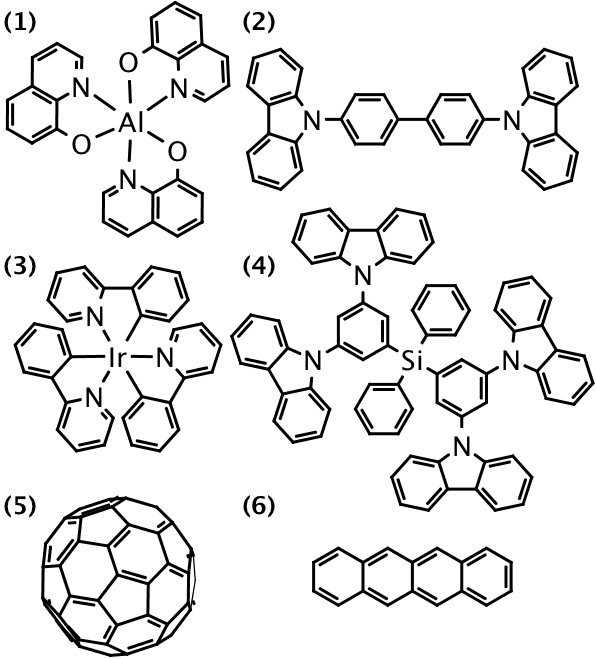}
\caption{Some example molecules successfully analyzed with atom probe tomography: (1) tris-(8-hydroxyquinoline)aluminum ($\mathrm{Alq_3}$), (2) 4,4'-bis(N-carbazolyl)-1,1'-biphenyl (CBP), (3) tris[2-phenylpyridinato-C2,N]iridium(III) (\irppy), (4) bis[3,5-di(9H-carbazol-9-yl)phenyl]diphenylsilane (SimCP2), (5) \full, (6) tetracene.}
\label{fig:materials}
\end{figure}

Organic molecular materials span a wide range of structures, bond types, masses, and atomic constituents. This variety makes organic electronics very versatile, but presents a challenge when trying to generalize their properties.
We have successfully analyzed a number of materials systems with APT that represent a reasonable cross-section of common organic electronic molecules including organometallics such as tris-(8-hydroxyquinoline)aluminum ($\mathrm{Alq_3}$) and tris[2-phenylpyridinato-C2,N]iridium(III) (\irppy), fullerenes, tetracene, and polycyclic aromatics such as 4,4'-bis(N-carbazolyl)-1,1'-biphenyl (CBP) and bis[3,5-di(9H-carbazol-9-yl)phenyl]diphenylsilane (SimCP2) as shown in Figure \ref{fig:materials}, and other materials we have successfully run include bathophenanthroline (BPhen), diindenoperylene (DIP), and 4-(dicyanomethylene)-2-methyl-6-julolidyl-9-enyl-4H-pyran (DCM2);
Table S1 of the Supporting Information provides information on various materials and their analysis parameters.
Thus far we have confined ourselves to thermally evaporable molecules, but this is not a known limitation of the technique; in fact, in electron irradiated systems we observe field evaporation of covalently bridged \full\ dimers and trimers (see Supporting Information Figure S5), which are both insoluble and cannot be sublimed. The limits for applicability of APT to organic and other materials systems are not fully known, and merit further study.

In applying a new technique to any system, it is important to understand the quality of information it provides. For APT, there are three major concerns:
(1) species discrimination,
(2) molecular fragmentation,
and (3) spatial resolution.
We address each of these to validate APT for organic molecular materials.

\subsection{Validation}
\begin{figure} % C60 isotope figure
\centering
\includegraphics[width=3.3in]{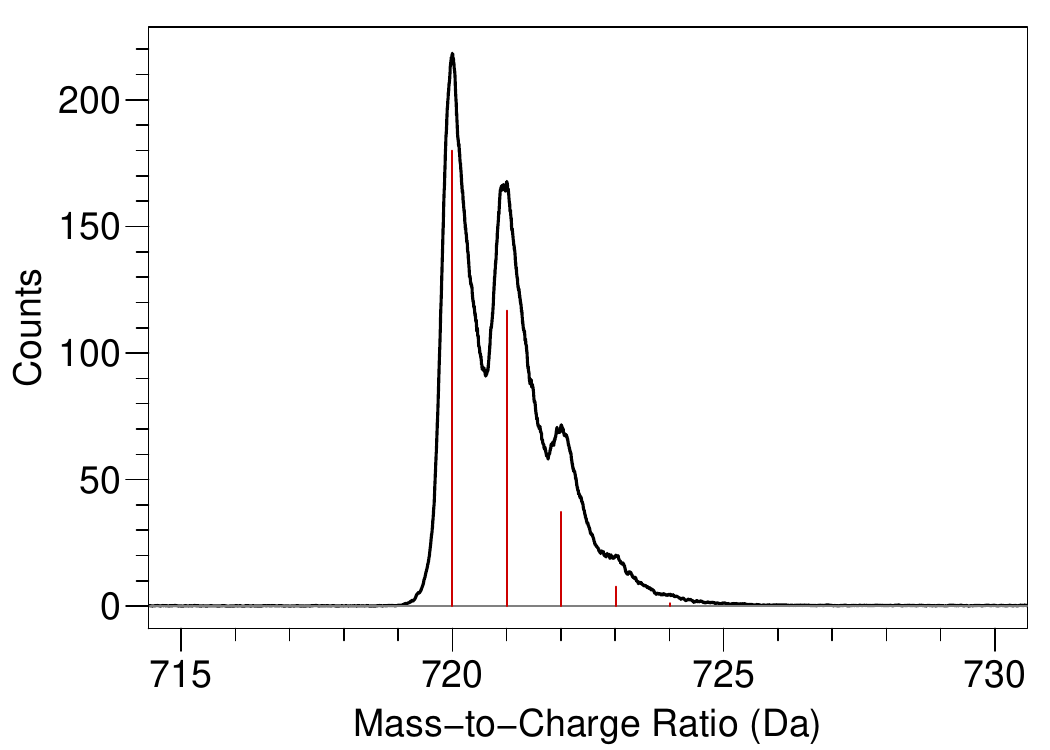}
\caption{Mass spectrum of $\mathrm{C_{60}^+}$ showing isotopic peaks; the red lines are its expected isotopic distribution. Based on the peak separation shown here, this spectrum has a mass resolving power ($m/\Delta m$) of about 1000.}
\label{fig:mrp}
\end{figure}

In APT, mass resolution is typically characterized by the mass resolving power (MRP), which is given by:%
\cite{Gault2012}
\begin{equation}
MRP = \frac{m}{\Delta m}
\end{equation}
where $\Delta m$ is the full-width at half-maximum of the peak at mass $m$. We have demonstrated MRPs of $> 1000$ using a flight path length of \SI{160}{\mm}. Figure \ref{fig:mrp} shows part of a mass spectrum collected on a sample of \full\ (structure shown in Figure \ref{fig:materials} (5)) in which the isotopic peaks are clearly resolved; based on the peaks shown, this this spectrum has a mass resolving power ($m/\Delta m$) of about 1000. This high MRP allows definitive identification of the molecular constituents of a sample by comparing the spectrum to the expected isotopic distribution.

\begin{figure*} % CBP impurity peak figure
\centering
\includegraphics[width=6in]{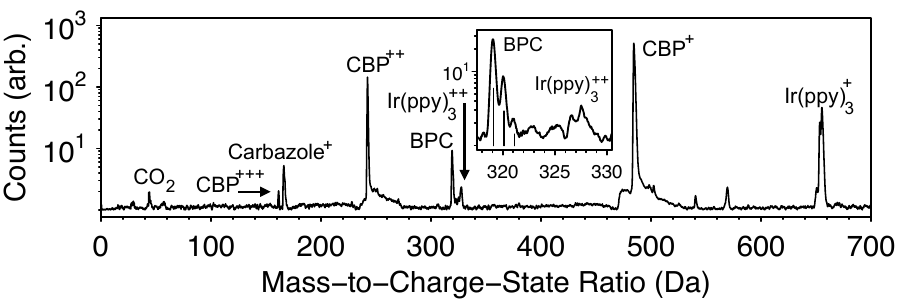}
\caption{Mass spectrum of a blended film of \SI{6}{vol.\percent} \irppy\ in purified CBP; (inset) Region of the impurity 4-N-(carbazolyl)biphenyl (BPC) showing a clear offset of \SI{1}{\dalton} from the expected fragment location, while the $\mathrm{Ir(ppy)_3^{++}}$ peak is at its expected mass; the vertical lines in the inset show the expected isotopic positions of the impurity.}
\label{fig:impurity}
\end{figure*}
Figure \ref{fig:impurity}a shows the mass spectrum of a blended film of \SI{6}{vol.\percent} \irppy\ in CBP (purified by thermal gradient sublimation) (their molecular structures are shown in Figure \ref{fig:materials} (2) \& (3)). There is a small but clearly resolvable peak at \SI{319}{\dalton} (marked BPC); that we can observe this peak at all is a clear advantage of APT in organic systems, as it comprises only
$\sim\SI{0.5}{\percent}$
of the total film. This peak is near the mass of CBP missing one carbazole group, and therefore might be thought to arise from the fragmentation of the molecule during the atom probe evaporation process; however, the high MRP of APT allows us to discern that this peak is \SI{1}{\dalton} too heavy for a fragment with a dangling bond, indicating that there is a hydrogen at the 4' position of the biphenyl group. This extra hydrogen suggests this peak is the impurity 4-(N-carbazolyl)biphenyl (BPC)%
\cite{Kondakov2007}
in the material left over from its synthesis or created during film deposition, not a fragment formed during the field evaporation process; we note that the nearby $\mathrm{Ir(ppy)_3^{++}}$ peak is at the expected mass, confirming that the transformation from TOF to mass-to-charge is accurate.
Therefore, if any fragmentation is occurring, such as the much smaller peaks in the spectrum, they comprise only a small fraction ($<\SI{1}{\percent}$) of the data. Furthermore, a run of a blended film of \SI{6}{vol.\percent} \irppy\ in \textit{as-received} CBP (Figure S6 in the Supporting Information) shows this impurity comprising about \SI{5}{\percent} of the film (with other small peaks similarly increased), further supporting the impurity interpretation of these peaks. Given the noise floor observed in these spectra, we estimate our sensitivity to be approximately \SI{50}{ppm}.

APT provides more data than just spatial position and mass, and this ancillary information can reveal more about what has been observed. Because of both the dynamic distribution of the electric field on the sample's surface and the stochastic nature of field evaporation events, multiple ions can evaporate during a single pulse.%
\cite{DeGeuser2007}
This is sometimes reduced through the selection of run parameters, but these multiple hit events can enhance our interpretation of the data. Two dimensional correlation histograms of double hit events during APT runs (Figures S7 \& S8 in the Supporting Information) show no evidence of molecular fragmentation during post-ionization of the \irppy:CBP sample, strengthening our rejection of significant fragmentation.%
\cite{Saxey2011}
We note that similar results are seen for other materials, such as $\mathrm{Alq_3}$ and even SimCP2, which is one of the larger evaporable small molecules (\ie\ \SI{977}{\dalton}).

Because the needle-like shape of the specimen acts as the optic and we use a much larger radius of curvature for our sample than typical APT specimens, it is critical to characterize our spatial resolution.
To test our spatial resolution, we prepared a film of \full\ on DIP, which templates \full\ with the (111) plane parallel to the substrate and enhances crystallinity;%
\cite{Hinderhofer2013}
an x-ray diffraction measurement of the film shows that it is textured as expected (Figure S9 in the Supporting Information).
The crystalline structure provides an internal measure of the spatial resolution of APT for our specimens, which has been the standard method of estimating resolution among the APT community.%
\cite{Gault2009a}

\begin{figure}% SDM of templated C60
\includegraphics[width=3.3in]{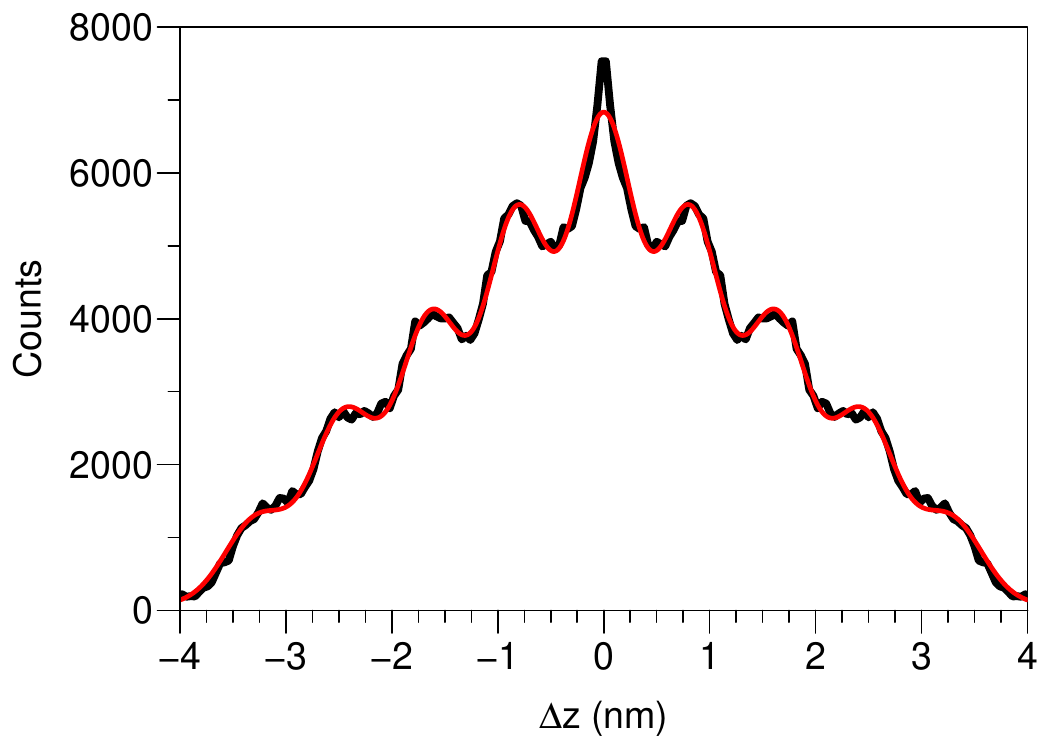}
\caption{%
Spatial distribution map (SDM) of \full\ templated on diidenoperylene (DIP) showing crystal lattice planes in the z direction. The sampled volume is $\sim\SI{2600}{\nm\cubed}$. A fit to the peaks (red) indicates the spatial resolution of this sample in the z dimension is $\sim\SI{0.3}{\nm}$.
}
\label{fig:SDM}
\end{figure}

Figure \ref{fig:SDM} shows the film's z-axis SDM---a common analysis method for crystalline APT data that measures the distribution of z-axis components of the vectors between points in a neighborhood.%
%; the components can be combined radially to yield the familiar radial distribution function (RDF).
\cite{Geiser2007}
We used a neighborhood of diameter \SI{2.4}{\mm} on the detector, corresponding to a sampled volume of $\sim\SI{2600}{\nm\cubed}$; formally, the resolution of the reconstruction is only well-defined in this region, but for the following discussion we extrapolate to the entire detected region.
Fitting to a model of evenly spaced Gaussian functions with equal standard deviations allows us to extract information about our resolution.
The regular spacing of the peaks is \SI{0.817(2)}{\nm} with this reconstruction using the model fit, which corresponds to the expected (111) lattice planes of the oriented \full\ at \SI{0.817}{\nm}.%
\cite{Fleming1992}
The width of the peaks indicates that the spatial resolution of this sample---which we define as the standard deviation of the fitted Gaussians---is \SI{0.313(3)}{\nm} in the z-direction, or $\sim\SI{0.3}{\nm}$ taking other uncertainties into account. We note that this is not a physical limit of the resolution for the technique, but what was achieved for this particular sample.

Given this known dimension in z, we can adjust the reconstruction parameters to return the expected density of the crystal. This reconstruction allows for estimation of the field of view, which for this sample is around \SI{760}{\nm}; as there are about 800 pixels across the detector,%
\cite{Larson2013}
our x-y resolution is limited to about \SI{1}{\nm} but could be improved with different sample geometries or increasing the detector path length. A real space plot of molecular locations and a three-dimensional reconstruction showing the crystal planes is provided in the Supporting Information (Figures S10 \& S11).

\subsection{Organic Electronics}
In 2016, we used APT to investigate the behavior of the bilayer \full/tetracene (Figure \ref{fig:materials} (5) \&\ (6)) small-molecule OPV system---one of the most studied models for testing our understanding of the crucial heterojunction interface for OPVs.%
\cite{Proudian2016}
APT revealed that the \full\ and tetracene molecules chemically react at the heterojunction interface, creating a partial monolayer of a Diels-Alder cycloadduct (DAc) species that cannot be avoided using standard deposition techniques. This spatially resolved chemical resolution helps explain changes in device performance due to the presence of the DAc.%
\cite{Shao2007,Proudian2016}

This problem takes advantage of APT's strengths: there is no elemental composition change for this reaction, and its concentration and location in the film are critically important to understanding its impact on device performance. The development of this analysis, and the explanation that resulted, was only possible because of the unique capabilities of APT. Measuring the clear structure-property relations of an organic system moves the field closer to closing the experiment-theory loop.

In fluorescent and phosphorescent systems such as OLEDs, high enough local concentrations of the emitter molecule can cause concentration quenching---an issue where a sufficient dopant concentration allows for rapid non-radiative relaxation to the ground state, reducing efficiency.
\cite{Drexage1977,Lutz1981}
%The first report of concentration quenching in OLEDs was by \citet{Tang1989} two years after their seminal paper describing efficient organic electroluminescence.
%\cite{Tang1987}
Additionally, changes in the overall concentration of the OLED emitter guest strongly affect rates of triplet-triplet annihilation and triplet-polaron quenching.%
\cite{Reineke2007,Reineke2009,Song2010}
These concentration effects impact both efficiency, peak emission wavelength, and lifetime, and are compounded for systems in which the guest emitter molecule aggregates.%
\cite{Reineke2009}

\begin{figure*}% CBP:Ir(ppy)3 (6%) clustering figure
\centering
\includegraphics[width=6in]{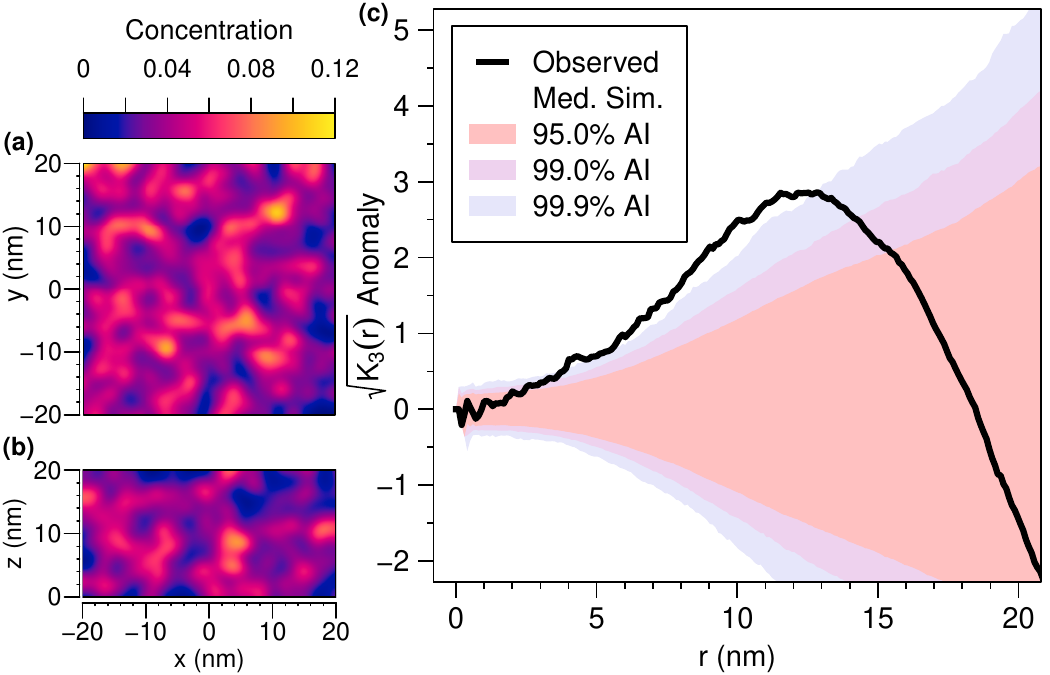}
\caption{Heat maps showing the concentration (fraction) of \irppy\ in a \SI[product-units = single]{40 x 40 x 20}{\nm} box projected onto \textit{(a)} the x-y plane and \textit{(b)} the x-z plane. \textit{(c)} K-function anomaly for \irppy\ in this box along with simulation envelopes. The excursion of the observed K-function above the envelopes indicates significant (deviation outside a \SI{99.9}{\percent} acceptance interval) clustering of the \irppy\ in the range of about \SIrange{5}{12}{\nm}.}
\label{fig:K}
\end{figure*}

The heavily-studied, archetypal phosphorescent OLED system of CBP:\irppy\ is suspected of this emitter aggregation, increasing self-quenching, exciton-exciton and exciton-polaron interaction, and device degradation.
\cite{Baldo1999a,Lamansky2001,Kawamura2006,Divayana2007,%
Reineke2009}
Measuring aggregation in OLED active layers is a natural problem for APT, which can easily discriminate and locate each molecule with high precision. To study aggregation, we deposited a blended film of 6 vol\% \irppy\ in CBP---a typical blend ratio for this system;%
\cite{Divayana2007}
the measured bulk composition of the blend is 5.8 ion\% \irppy;
we note that while these percentages differ in their calculation (vol\% versus ion\%), the molecules are close in size and the variation is within the error of the crystal monitor calibration.
We provide a region of the reconstruction in the Supporting Information (Figure S12).
A common method of looking for clusters is to look at heat map projections (shown in Figure \ref{fig:K}a,b) through the sample for ``hot spots;'' however, this does not indicate the significance of the clustering, as it relies on subjectively interpreting the data.

To formally characterize clustering of \irppy\ in the sample, we used the three-dimensional K-function ($K_3$); this function measures the number of additional \irppy\ molecules contained within a sphere of radius $r$ centered about each \irppy\ molecule, normalized by the bulk concentration of \irppy.%
\cite{Ripley1977}
Acceptance interval envelopes were generated for the measured K-function by randomly relabeling the molecular identities of the measured CBP:\irppy\ locations 50,000 times and calculating $K_3(r)$ of the relabeled \irppy\ patterns. Deviations above this envelope indicate significant non-random clustering of the observed \irppy\ at those distances.

We use the square root of the K-function to stabilize the variance of the envelopes; this transformation does not change any of the underlying conclusions drawn about the observed data, and is commonly used in the analysis of point patterns.
\cite{Ripley1977,Baddeley2016}
To make interpretation of the data easier, we also subtract the median value of the envelope, centering the plot about zero. From the transformed K-function and envelopes shown in Figure \ref{fig:K}c, we observe significant (deviation outside a \SI{99.9}{\percent} acceptance interval) clustering of the \irppy\ in the range of about \SIrange{5}{12}{\nm}. This corroborates earlier work using HAADF-STEM that suggested clustering of \irppy\ in 4,4',4''-tris(N-carbazolyl)-triphenylamine  (TCTA),%
\cite{Reineke2009}
but adds quantitative length scales in three dimensions that allow for better models of mixing and charge transport. Furthermore, this K-function analysis of APT data can be extended to general small-molecule organic systems, as it does not require a high z-contrast atom such as the Ir core in \irppy. The effect of the observed clustering on device properties will be explored in future publications.

\section{Conclusions}
Morphology is inextricably linked to the behavior of organic devices, and it is crucial to understand---and ultimately control---the structures of these systems. Using APT to study small-molecule organic semiconducting systems can potentially answer long-standing questions of device characteristics, and bolster a broader theory of organic electronic physics. The high chemical discrimination ($<\SI{1}{\dalton}$), spatial resolution ($\sim\SI{0.3}{\nm}$ in z and $\sim\SI{1}{\nm}$ in x-y), and analytic sensitivity ($\sim\SI{50}{ppm}$) we have achieved with APT are unmatched for studying the morphology of small-molecule organic materials; we note that these parameters are not fully optimized and there is still room for improvement.
Some work has been done to use a FIB to lift out a section of organic material from a film.%
\cite{Kim2011,Orthacker2014}
If low-damage FIB lift-out of these films can be developed, it opens up many more avenues for experimentation. In particular, it permits observations of the morphological differences between devices before and after operation for extended times, and may allow for improved x-y resolution. The spatial information provided for working devices is particularly valuable for closing the experiment-theory loop.

Through chemically-specific and spatially-resolved information, APT can answer questions about both bulk and interfacial structures. This capability adds another powerful characterization tool for the organic electronics community, enabling new experimental directions. As APT matures, it will become an ever more valuable technique for studying organic molecular materials. By providing high-resolution and chemically-specific information, APT can help move organic semiconductors towards their promise of large-scale, high-versatility, and low-cost electronics.

\section{Acknowledgments}
Work by APP, MBJ, and JDZ was supported by the U.S. Department of Energy, Office of Science, Basic Energy Sciences under Award DE-SC0018021. The LEAP was supported by a National Science Foundation Major Research Instrumentation grant DMR-1040456.

\section{Supporting Information}
The supporting information provides:
an SEM image of a Si sample tip,
a cross-sectional plot of the reconstructed lenticular shape of our samples,
a representative detector hit map and voltage curve from APT,
a mass spectrum of electron irradiated \full\ showing dimers and trimers compared to that of a pristine, unirradiated \full\ sample,
a mass spectrum of as-received \irppy:CBP showing increased levels of the impurity peak,
correlation histograms showing no fragmentation curves for the purified  and as-received \irppy:CBP samples,
an XRD scan of a witness for the SDM sample film, 
a real-space plot of the points used for the SDM with visible lattice planes,
a three-dimensional reconstruction of the volume used for the SDM showing regions of high density,
a three-dimensional reconstruction of the volume used for the clustering analysis,
and a table of materials analyzed with APT process parameters.

\bibliography{biblio}
%\pagebreak
%\centering{\includegraphics[width=3.3in]{toc}}
\end{document}

\vspace{-1.5em}
\maketitle

\renewcommand{\thefigure}{S\arabic{figure}}
\setcounter{figure}{0}
\renewcommand{\thetable}{S\arabic{table}}
\setcounter{table}{0}

\section{Atom Probe}
\begin{figure}
\includegraphics[width=\linewidth]{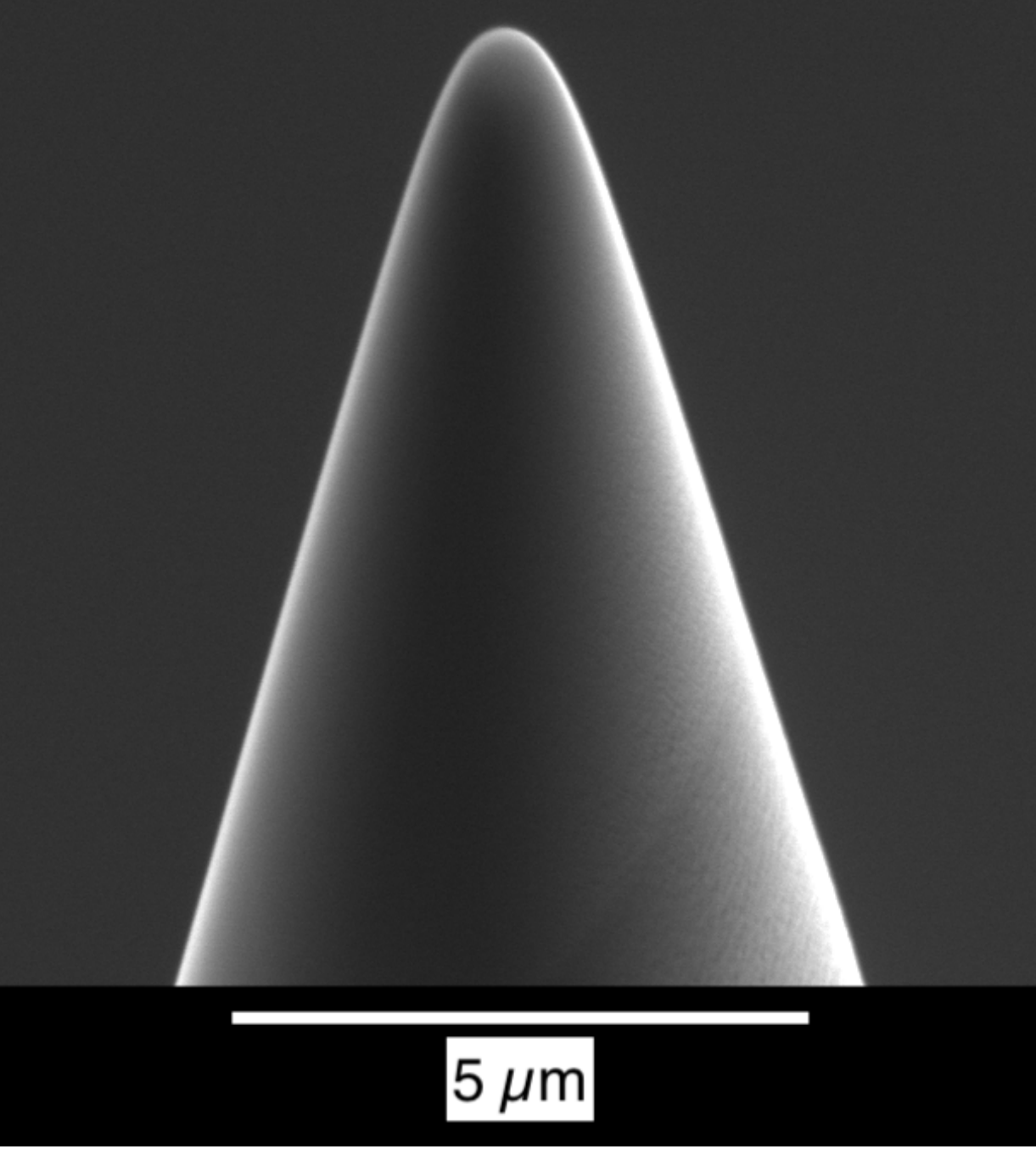}
\caption{A representative Si tip used for film deposition and subsequent atom probe tomography (APT) analysis.}
\label{fig:tip}
\end{figure}

\begin{figure}
\includegraphics[width=\linewidth]{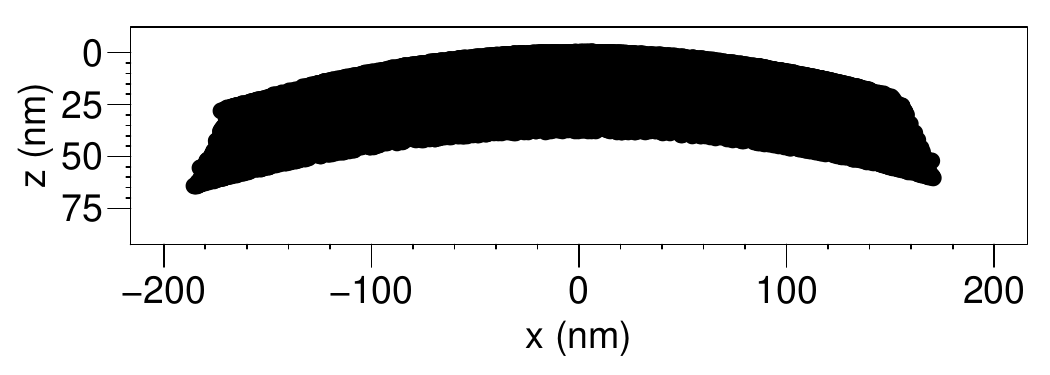}
\caption{A cross-section of a \full\ film reconstruction which illustrates the lenticular shape of the analyzed film typical of this sample preparation method.}
\label{fig:film}
\end{figure}

\begin{figure}
\includegraphics[width=\linewidth]{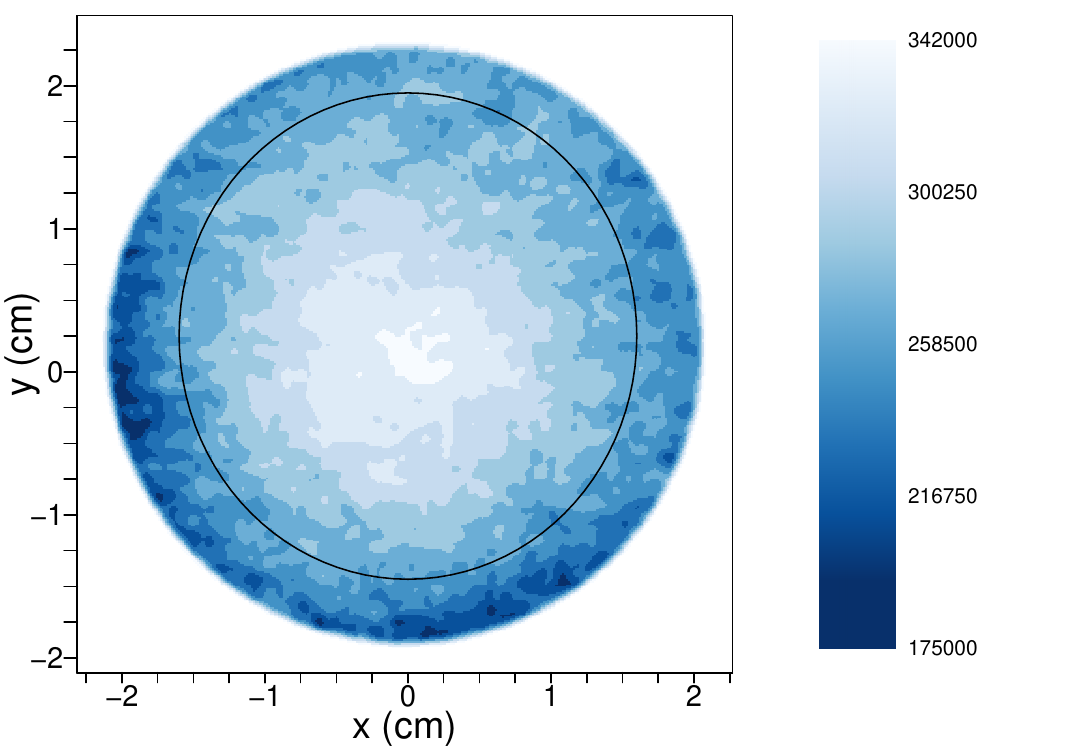}
\caption{A representative detector hit map for small-molecule organic semiconductor samples.}
\label{fig:det}
\end{figure}

\begin{figure}
\includegraphics[width=\linewidth]{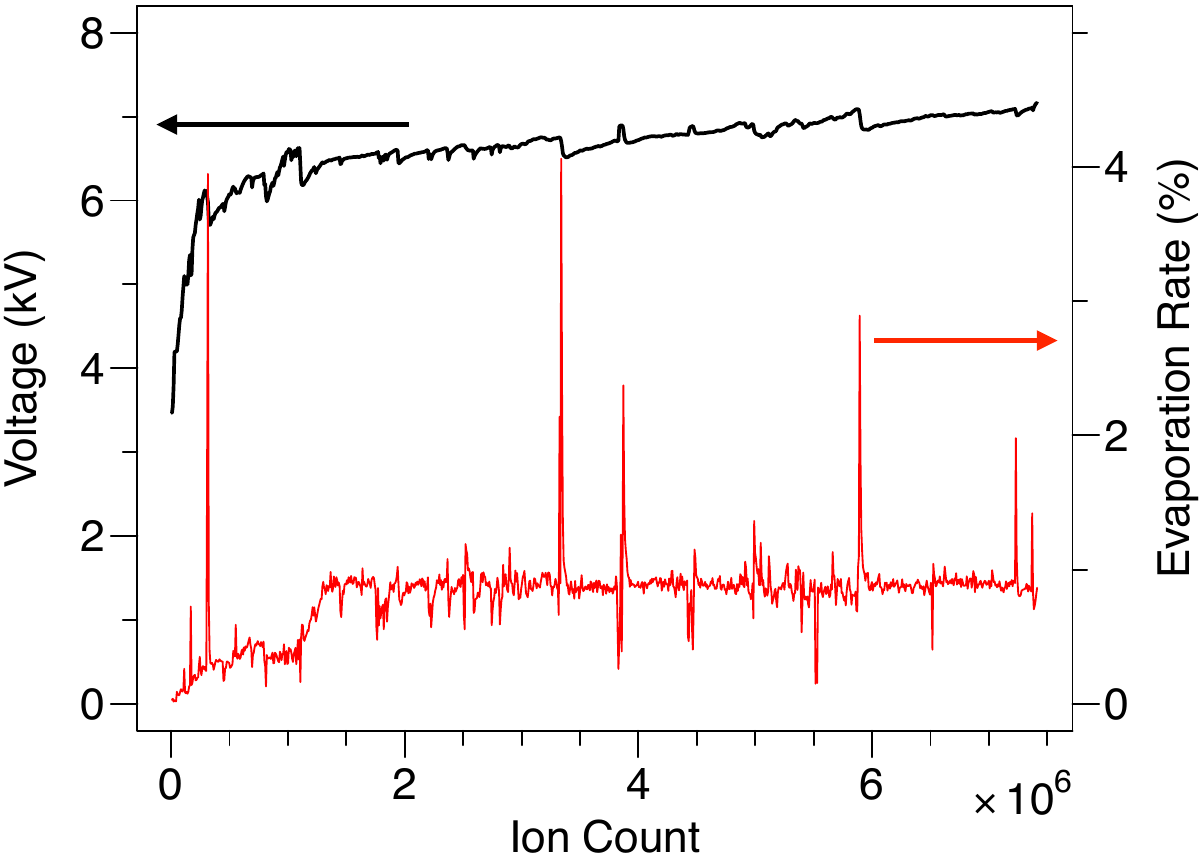}
\caption{A representative voltage curve (black) and corresponding evaporation rate (red) for small-molecule organic semiconductor samples. The fluctuations in the curve are due to the laser drift compensation algorithm not working effectively for our sample geometry, requiring manual laser adjustments to keep the sample properly running.}
\label{fig:volt}
\end{figure}

We use a smooth Si tip as the substrate for our APT sample films (Figure \ref{fig:tip}). Because of their curvature and the deposition method used to create samples, the analyzed films have a lenticular shape; we illustrate the shape of the films in Figure \ref{fig:film}. Though very large compared to a typical APT sample radius of curvature,%
\cite{Kelly2012}
these tips still yield smooth evaporation both spatially and temporally. Figure \ref{fig:det} shows the smooth variation of hits across the detector, and Figure \ref{fig:volt} shows the narrow window of voltage used during APT. We note that the fluctuations in voltage during the run are due to laser drift compensation algorithm errors arising from the much larger radius of our tip, requiring manual laser adjustments to keep the sample properly running; this leads to more abrupt changes in evaporation rate and hence voltage than in typical APT runs.


\section{Electron Irradiation}
\begin{figure}[h]
\includegraphics[width=\linewidth]{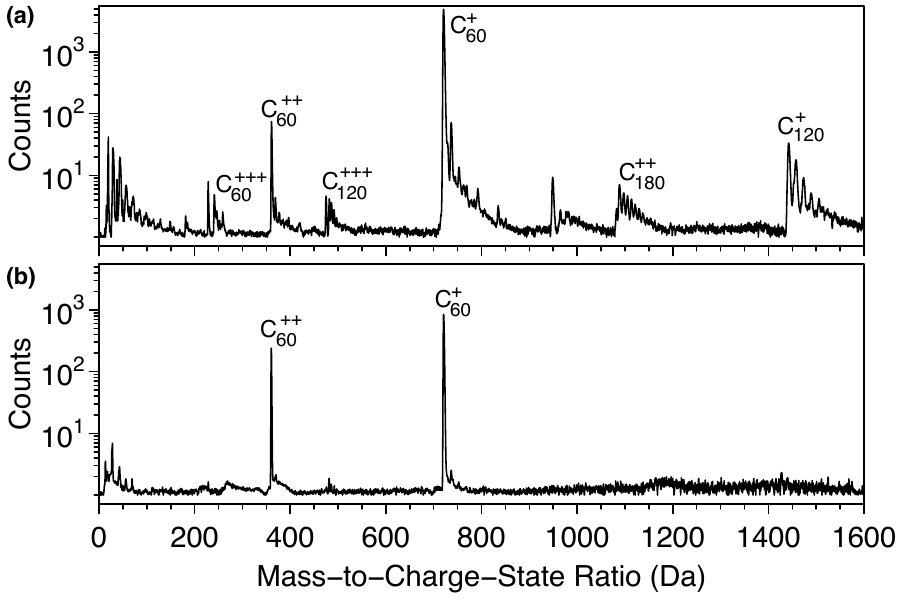}
\caption{\textit{(a)} Mass spectrum of electron-irradiated \full, showing evaporation of dimers (singly ionized peaks at \SI{1440}{\dalton} and triply ionized peaks at \SI{480}{\dalton}; the doubly ionized peaks are subsumed in the primary \full\ peak) and trimers (doubly ionized peaks starting at \SI{1080}{\dalton}) of \full. The peaks labeled Tc and DAc are tetracene and a Diels-Alder adduct of \full\ and tetracene; they are present because of background tetracene contamination in the deposition chamber.
\textit{(b)} Mass spectrum of unirradiated \full\ showing only \full\ peaks with no dimers or trimers.}
\label{fig:e_C60}
\end{figure}

Because small-molecule organic films are sensitive to electron irradiation, the common atom probe practice of imaging the sample in an SEM or TEM prior to performing APT can damage the film. Figure \ref{fig:e_C60}a shows the mass spectrum of a \full\ sample that was imaged in an SEM before analysis, which created dimers and trimers of \full. These species are insoluble and cannot be sublimed, meaning that they must have been created during the electron imaging process. That they evaporate in during APT suggests that materials may be analyzed using APT that are not necessarily thermally evaporable, given the right processing conditions. We note that these \full\ dimers and trimers are not observed in appreciable quantities in films that have not undergone SEM imaging (Figure \ref{fig:e_C60}b) under similar analysis conditions.

\section{Impurity}
\begin{figure}[h]
\includegraphics[width=\linewidth]{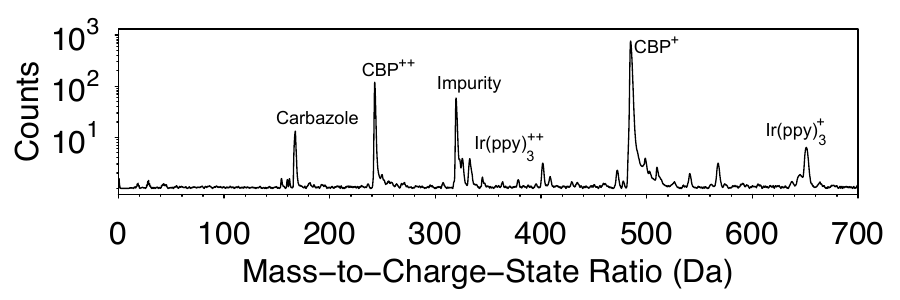}
\caption{Mass spectrum of \SI{6}{vol.\percent} \irppy\ in \textit{as-received} CBP; the impurity at \SI{319}{\dalton} is \SI{5}{\percent} as compared to \SI{0.5}{\percent} in the CBP purified by thermal gradient sublimation used in the \SI{6}{vol\percent} \irppy:CBP blend (Figure 3).}
\label{fig:cbp_ms}
\end{figure}

\begin{figure}[h]
\includegraphics[width=\linewidth]{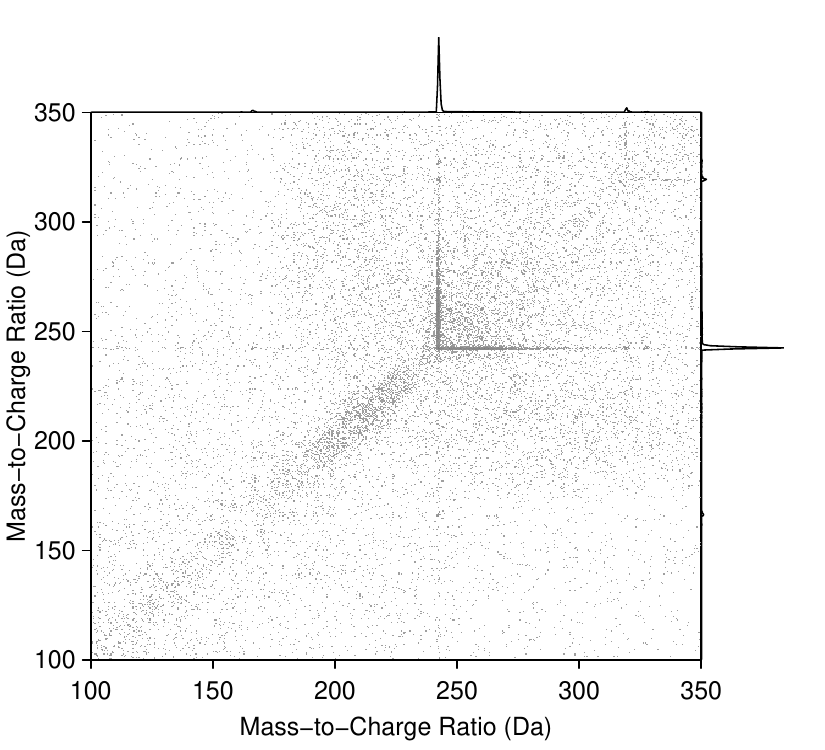}
\caption{A correlation histogram of a blended film of \SI{6}{vol.\percent} tris[2-phenylpyridinato-C2,N]iridium(III) (\irppy) in 4,4'-bis(N-carbazolyl)-1,1'-biphenyl (CBP, purified by thermal gradient sublimation) focused on the $\mathrm{CBP^{++}}$ peak, which shows no evidence of fragmentation of the CBP into the unexpected peaks.}
\label{fig:oled_sax}
\end{figure}

\begin{figure}[h]
\includegraphics[width=\linewidth]{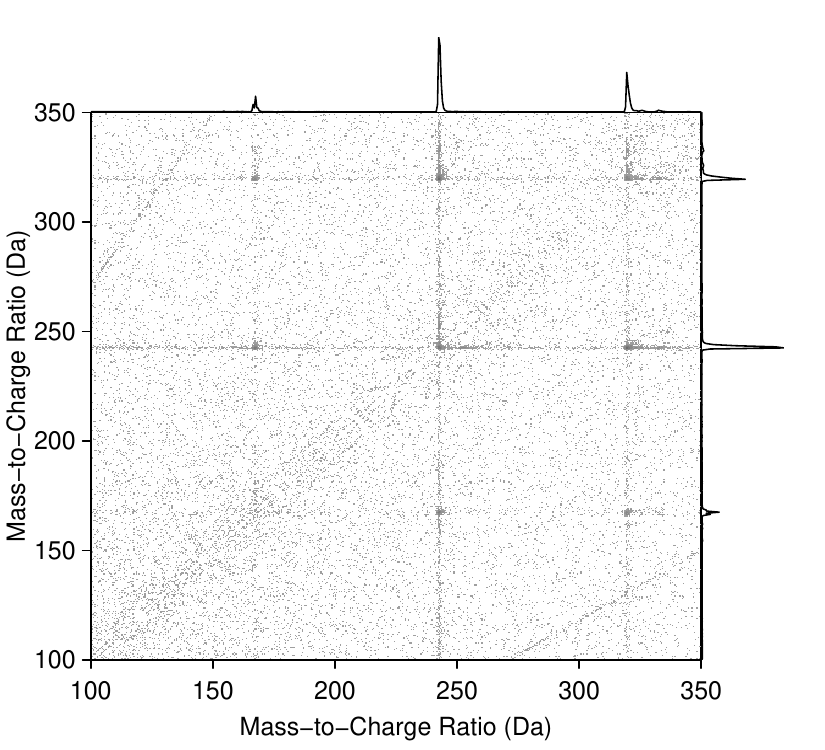}
\caption{A correlation histogram of the as-received CBP focused on the $\mathrm{CBP^{++}}$ peak at \SI{242}{\dalton}, which shows no evidence of fragmentation of the CBP into the unexpected peaks.}
\label{fig:cbp_sax}
\end{figure}

For the blended film of \SI{6}{vol.\percent} tris[2-phenylpyridinato-C2,N]iridium(III) (\irppy) in 4,4'-bis(N-carbazolyl)-1,1'-biphenyl (CBP, purified by thermal gradient sublimation), there are three pieces of evidence that support assignment of the peak at \SI{319}{\dalton} as an impurity over a fragment. First is the mass detected in the inset of Figure 3 indicating the presence of H at the possible fragmentation location. Second is a correlation histogram of the data showing no evidence of fragmentation of the doubly ionized CBP into that peak (Figure \ref{fig:oled_sax}). Third is the relative component of the peak is an order of magnitude higher in the as-received material (\SI{5}{\percent}) (Figure \ref{fig:cbp_ms}) as compared to the purified (\SI{0.5}{\percent}) (Figure 3). In the film with as-received CBP, with much a higher signal from this impurity peak, there is still no evidence of fragmentation in a correlation histogram (Figure \ref{fig:cbp_sax}).

Correlation histograms of events when two ions are detected for a single laser pulse were proposed by \citeauthor{Saxey2011} to aid in mass spectrum analysis for atom probe tomography (APT).%
\cite{Saxey2011}
It plots the masses ($i$ \& $j$) of each detection event against each other and then bins this data to examine occurrence frequencies; as a result, it is symmetric about $i=j$.

There are four key features that are visible in correlation histograms. First are the expected bright spots at coincidences between major peaks $i$ and $j$ representing the field evaporation of two ions. More interesting are the three types of lines that emanate from these points. Horizontal and vertical lines are due to the delayed evaporation of one ion of the pair. Curved tracks that go from low $i$ and $j$ (lower left) to high $i$ and $j$ (upper right) are the result of delayed evaporation for both ions. Finally, tracks that go from upper left to lower right are due to mid-flight dissociation of the parent ion, which are of interest when investigating possible molecular fragmentation. We note that no tracks associated with mid-flight dissociation are observed in the data.


\section{Reconstruction}
To verify that our reconstruction validation sample (Figures 4, \ref{fig:real} \&\ \ref{fig:conc}) was crystallized with (111) texture, we performed XRD on a co-deposited witness sample. 
XRD data was collected on a
Panalytical PW3040 X-ray diffractometer
using Cu-$\mathrm{K_\alpha}$ radiation in the
Bragg-Brentano geometry with five Soller slits on the incident and receiving sides over
\SIrange{5}{30}{\degree} using a \SI{0.01}{\degree} step with \SI{5}{\second} integration.
The peaks at \SIlist{10.8;21.7}{\degree} correspond to an inter-planar spacing of \SI{0.817}{\nm}---matching the (111) plane spacing in \full---and no other peaks are present, which suggests that the \full\ is indeed (111) textured.
The peak intensity is much higher than in non-templated \full, suggesting that the crystallinity of the \full\ is enhanced, as demonstrated by \citet{Hinderhofer2013}

\begin{figure}
\includegraphics[width=\linewidth]{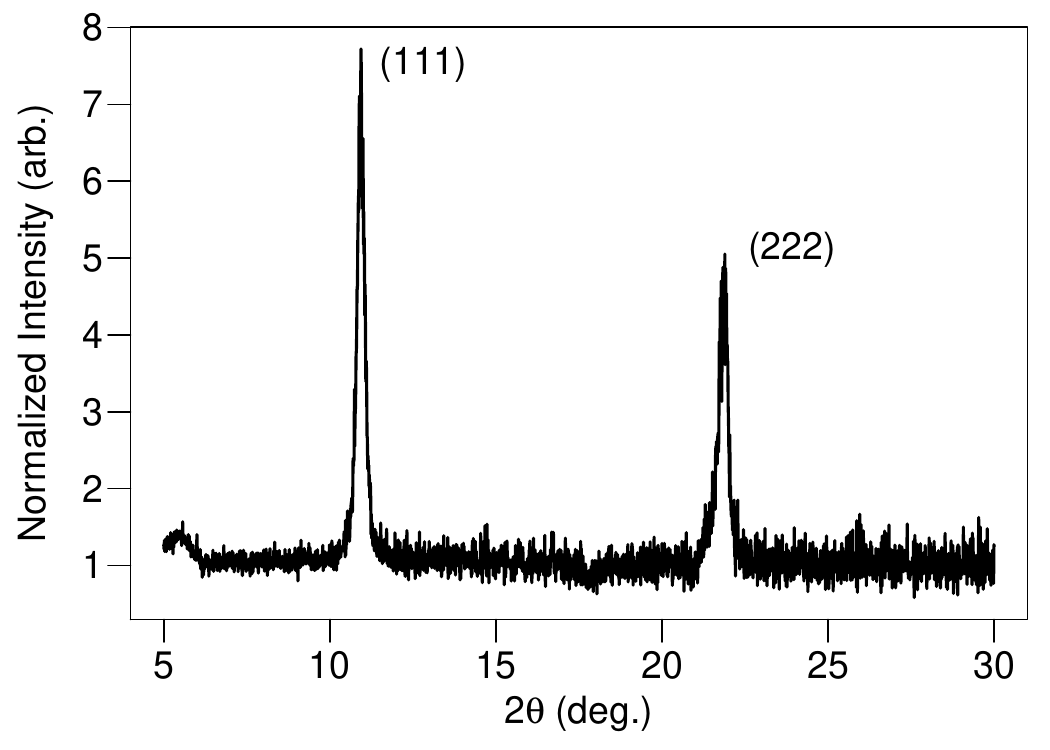}
\caption{An x-ray diffraction (Cu-$\mathrm{K_\alpha}$) measurement of the diidenoperylene/\full\ film showing it is textured with the (111) plane parallel to the substrate.}
\label{fig:xrd}
\end{figure}
\begin{figure}
\includegraphics[width=\linewidth]{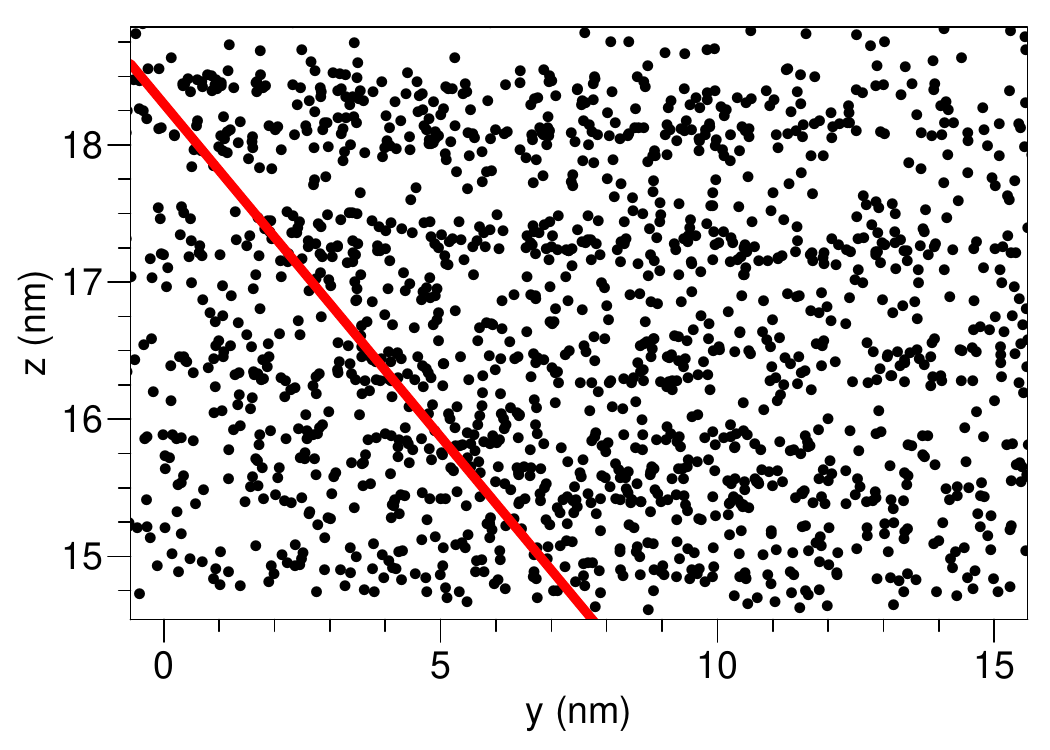}
\caption{A real space plot of the points used to generate the spatial distribution map (SDM) in Figure 4. The red line marks the location of a potential grain boundary.}
\label{fig:real}
\end{figure}
\begin{figure}
\includegraphics[width=\linewidth]{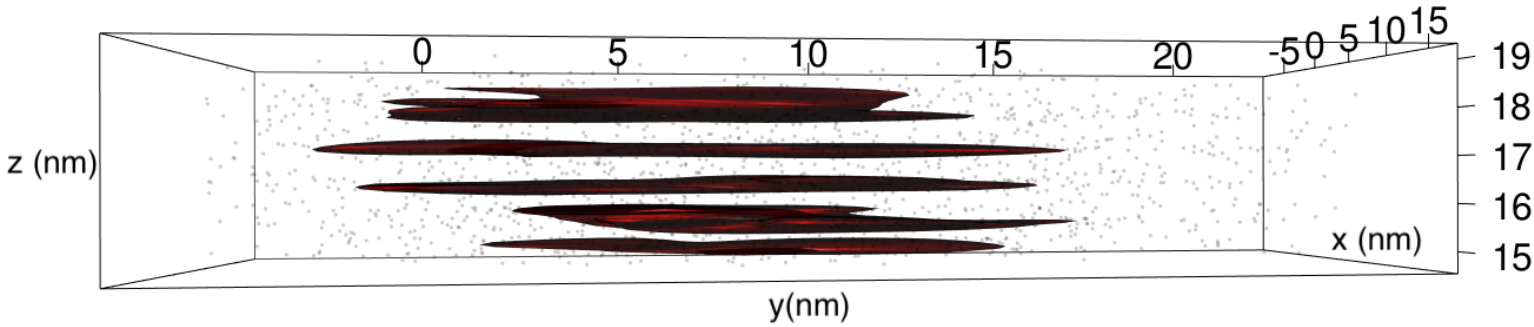}
\caption{A three-dimensional plot of the points used to generate the spatial distribution map (SDM) in Figure 4 showing the high concentration regions of the crystal planes.}
\label{fig:conc}
\end{figure}

\section{Clustering}
\begin{figure}
\includegraphics[height=8in]{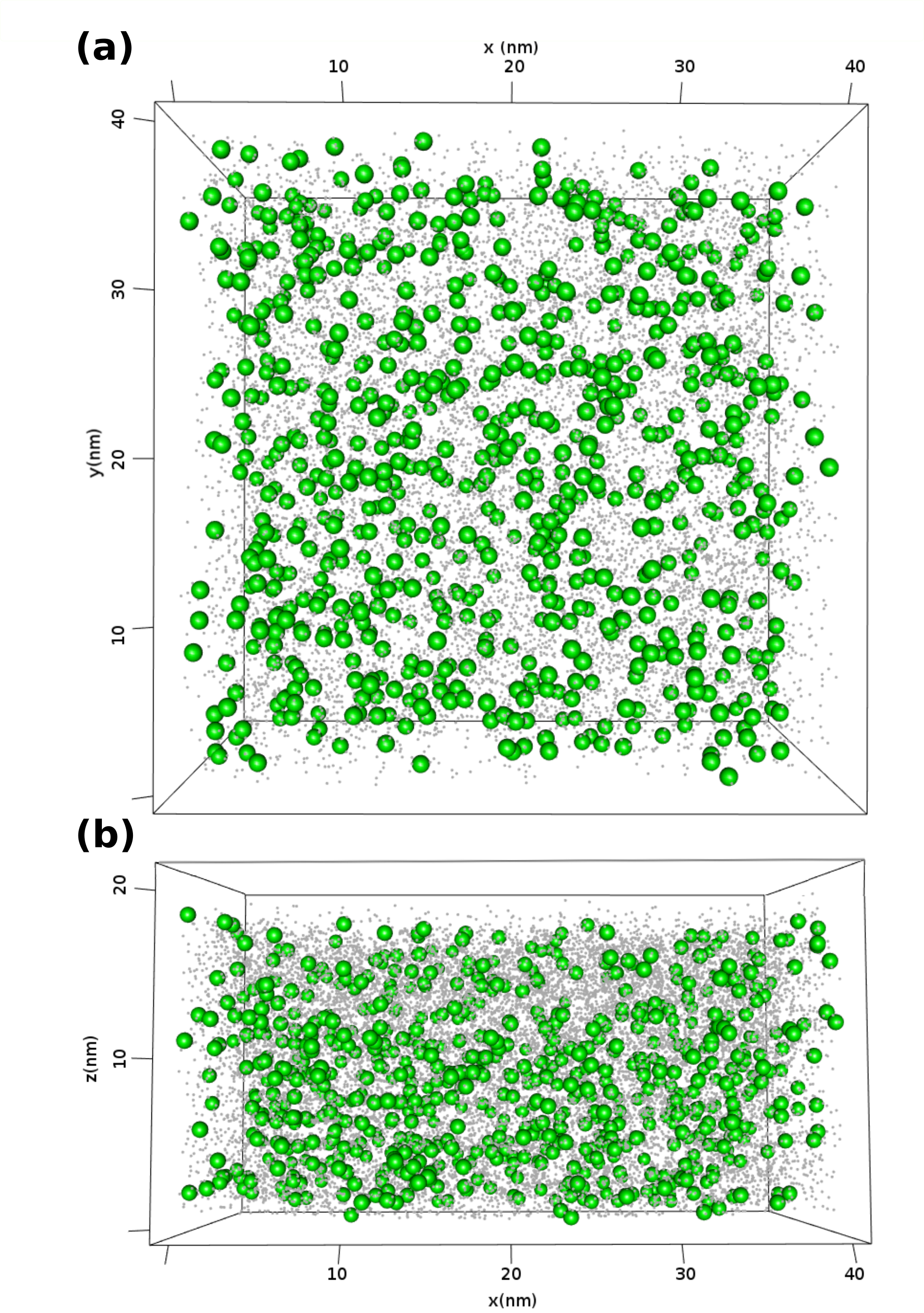}
\caption{A three-dimensional plot of the points used to generate the $K_3$ cluster analysis shown in Figure 5c from the perspective of the heat maps in Figure 5a--b. The black dots are CBP while the green spheres are \irppy.}
\label{fig:cluster}
\end{figure}

For the \irppy:CBP sample, we characterize clustering using the $K_3$ function.%
\cite{Ripley1977}
This is necessary because inspecting the film by eye does not readily distinguish lower levels of dopant clustering. Figure \ref{fig:cluster} shows the region of the film used for cluster analysis from the two perspectives of the heat maps shown in Figure 5a--b, demonstrating the utility of the heat maps and K-function analysis.

\section{Material Evaporation}
Sample run parameters and the corresponding voltage at which the signal became apparent are shown in Table \ref{tab:evap}. The tip radius and the ``turn-on'' voltage are estimates. In particular, the starred radii are from W tips instead of Si, which---due to their fabrication process---are much more variable about their nominal radius.

\begin{table}
\begin{tabular}{ccccc}
\textbf{Material} & \textbf{Radius (nm)} & \textbf{Temperature (K)} & \textbf{Energy (pJ)} & \textbf{Voltage (kV)} \\ \hline
\alq   & 250  & 30 & 15 & 3.8 \\
\alq   & 500* & 40 & 20 & 3.9 \\
BPhen  & 250  & 40 & 30 & 1.7 \\
\full  & 250  & 35 & 12 & 2.2 \\
\full  & 500* & 40 & 60 & 1.2 \\
CBP    & 250  & 30 & 8  & 3.5 \\
DCM2   & 500* & 40 & 20 & 4.3 \\
DIP    & 250  & 35 & 12 & 3.0 \\
DIP    & 500  & 30 & 12 & 7.5 \\
\irppy & 250  & 40 & 30 & 2.5 \\
\irppy & 500  & 30 & 18 & 6.2 \\
mCBP   & 250  & 40 & 30 & 2.3 \\
SimCP2 & 250  & 30 & 10 & 3.9 \\
Tc     & 500* & 40 & 20 & 2.2 \\
Tc     & 500* & 40 & 50 & 1.4 \\
TCP    & 250  & 30 & 10 & 3.5 \\
TCP    & 500  & 30 & 10 & 7.0
\end{tabular}
\caption{A table of evaporation parameters for a variety of materials that we have successfully run in the atom probe. The tip radii are nominal values; the stars indicate that the tip was made from W instead of Si, and have a larger uncertainty in their radius because of their fabrication process. The ``turn-on'' voltage was estimated from the voltage at which the species first became apparent in the mass spectrum.}
\label{tab:evap}
\end{table}

%The composition measurements are consistent with expectations as well. Table \ref{tab:conc} shows the expected dopant concentration vs. measured concentration in the atom probe, showing good agreement to within the error of deposition.
%
%\begin{table}
%\begin{tabular}{ccc}
%\textbf{Material} & \textbf{Nominal Conc. (vol. \%)} & \textbf{Measured Conc. (mol. \%)} \\ \hline
%\irppy\ in CBP    &  6 & 5.8 \\
%\irppy\ in CBP    & 10 & \\
%\irppy\ in mCBP   & 10 & \\
%\irppy\ in SimCP2 &  6 & \\
%\irppy\ in TCP    & 10 & \\
%DCM2 in \alq      &  1 & \\
%DCM2 in \alq      &  2 & \\
%\end{tabular}
%\caption{A table the nominal versus expected dopant concentration for a variety of blended films.}
%\label{tab:conc}
%\end{table}

\bibliography{biblio}